\documentclass[12pt]{article}
\usepackage[dvips]{color}
\usepackage{epsfig}
\usepackage{amsmath}
\usepackage{graphicx}

\begin{document}
\title{\bf The center of mass energy of two colliding particles in the
STU black holes}
\author{{H. Saadat\thanks{Email: hsaadat2002@yahoo.com}}\\
{\small {\em  Department of  physics, Shiraz Branch, Islamic Azad
University, Shiraz, Iran}}\\
{\small {\em P. O. Box 71555-477, Iran}}} \maketitle
\begin{abstract}
In this paper we consider collision of two particle near the STU black hole and calculate center of mass energy.
In the case of uncharged black hole we find that the maximum energy obtained near the black hole horizon which similarly happen for charged black hole.
We verify that the black hole charge may be decreased or increased the center of mass energy near the black hole horizon.
\end{abstract}
\section{Introduction}
In the Ref. [1] the center-of-mass (CM) energy of a Kerr black hole
studied and proposed that this might lead to signals from ultra high
energy collisions such as dark matter physics [2, 3]. It yields to
universal property of rotating black holes which is infinite CM
energy of colliding particles and investigated by several papers
[4-16].\\
Now, aim of this paper is studying the particle acceleration
mechanism of STU black hole [16-22]. The STU black hole exist in
special case of $D=5$, $\mathcal{N}=2$ gauged supergravity theory
which is dual to the $\mathcal{N}=4$ SYM theory with finite chemical
potential. In this background, generally there are three electric
charges. In this paper we assume that three of them will be equal
and discussed about uncharged black holes. Therefore, first we
introduce STU black hole and then try to obtain CM energy of two
colliding particles near the STU black hole and then discuss about
consequence.

\section{STU black hole}
STU black hole is described by the following solution,
\begin{equation}\label{s1}
ds^{2}=-\frac{f}{{\mathcal{H}}^{\frac{2}{3}}}dt^{2}
+{\mathcal{H}}^{\frac{1}{3}}(\frac{dr^{2}}{f}+\frac{r^{2}}{R^{2}}d\Omega^{2}),
\end{equation}
where,
\begin{eqnarray}\label{s2}
f&=&1-\frac{\mu}{r^{2}}+\frac{r^{2}}{R^{2}}{\mathcal{H}},\nonumber\\
{\mathcal{H}}&=&\prod_{i=1}^{3} H_{i},\nonumber\\
H_{i}&=&1+\frac{q_{i}}{r^{2}}, \hspace{10mm} i=1, 2, 3,
\end{eqnarray}
and $R$ is the constant AdS radius which relates to the coupling
constant via $R=1/g$, and $r$ is the radial coordinate along the
black hole. The black hole horizon specified by $r=r_{h}$ which is
obtained from $f=0$. Also $\mu$ is called non-extremality parameter.
So, for the extremal limit one can assume $\mu=0$. Moreover
$d\Omega^{2}$ includes $d\theta^{2}$ and $d\phi^{2}$. The Hawking
temperature of STU black hole is given by,
\begin{equation}\label{s3}
T=\frac{r_{h}}{2\pi
R^{2}}\frac{2+\frac{1}{r_{h}^{2}}\sum_{i=1}^{3}{q_{i}}-\frac{1}{r_{h}^{6}}\prod_{i=1}^{3}{q_{i}}}{{\sqrt{\prod_{i=1}^{3}(1+\frac{q_{i}}{r_{h}^{2}})}}}.
\end{equation}
So, in the case of $q_{i}=0$ we get,
\begin{equation}\label{s4}
r_{h}=\pi R^{2}T.
\end{equation}
Here it is useful to discuss horizon structure of the metric (1).
The $f=0$ reduced to the following equation,
\begin{equation}\label{s5}
r^{6}+\mathcal{A}r^{4}-\mathcal{B}r^{2}+q_{1}q_{2}q_{3}=0,
\end{equation}
where,
\begin{equation}\label{s6}
\mathcal{A}\equiv q_{1}+q_{2}+q_{3}+R^{2},
\end{equation}
and,
\begin{equation}\label{s7}
\mathcal{B}\equiv \mu R^{2}-q_{1}q_{2}-q_{2}q_{3}-q_{1}q_{3}.
\end{equation}
A possible solutions of the equation (5) is given by,
\begin{equation}\label{s8}
r_{\pm}=\pm\left(\frac{W^{2}-2\mathcal{A}W+4(3\mathcal{B}+\mathcal{A}^{2})}{6W}\right)^\frac{1}{2},
\end{equation}
where,
\begin{eqnarray}\label{s9}
W^{3}&=&-36\mathcal{A}\mathcal{B}-108\prod_{i=1}^{3}q_{i}-8\mathcal{A}^{3}\nonumber\\
&+&12\sqrt{-12\mathcal{B}^{3}-3\mathcal{A}^{2}\mathcal{B}^{2}+54\mathcal{A}\mathcal{B}\prod_{i=1}^{3}q_{i}+81(\prod_{i=1}^{3}q_{i})^{2}+12\mathcal{A}^{3}\prod_{i=1}^{3}
q_{i}}.
\end{eqnarray}
The $r_{+}$ and $r_{-}$ denote outer and inner horizons
respectively.

\section{Center of mass energy}
In order to obtain the CM energy we consider planar motion which
yields $\dot{\theta}=0$. Other 4-velocity of the particles are
$\dot{t}$, $\dot{\phi}$ and $\dot{r}$. The first two components
obtained by using the following relations,
\begin{equation}\label{s10}
E=\frac{f}{{\mathcal{H}}^{2/3}}\dot{t},
\end{equation}
and,
\begin{equation}\label{s11}
L=\frac{r^{2}}{R^{2}}{\mathcal{H}}^{2/3}\dot{\phi},
\end{equation}
where $E$ and $L$ denote the test particle energy and the angular
momentum parallel to the symmetry axis per unit mass respectively.
In order to obtain $\dot{r}$ we assume,
\begin{equation}\label{s12}
S=\frac{1}{2}\tau-Et+L\phi+S_{r}(r),
\end{equation}
where $\tau$ is proper time. Then we use,
\begin{equation}\label{s13}
\frac{\partial S}{\partial\tau}=-\frac{1}{2}g^{\mu\nu}\frac{\partial
S}{\partial x^{\mu}}\frac{\partial S}{\partial x^{\nu}},
\end{equation}
and,
\begin{equation}\label{s14}
\frac{d S_{r}(r)}{\partial r}=g_{rr}\dot{r}.
\end{equation}
Therefore, we can summarize,
\begin{eqnarray}\label{s15}
\dot{t}&=&\frac{{\mathcal{H}}^{2/3}}{f}E,\nonumber\\
\dot{\phi}&=&\frac{R^{2}}{r^{2}}{\mathcal{H}}^{-1/3}L,\nonumber\\
\dot{r}&=&\sqrt{\frac{f}{{\mathcal{H}}^{1/3}}(1+\frac{{\mathcal{H}}^{2/3}}{f}E^{2}-\frac{R^{2}}{r^{2}}{\mathcal{H}}^{-1/3}L^{2})}.
\end{eqnarray}
If we take the angular momentum per unit mass $L_1$, $L_2$ and
energy per unit mass $E_1$, $E_2$, respectively, also $m_0$ as the
rest mass of both particles, then using the following relation,
\begin{equation}\label{s16}
E_{CM}=\sqrt{2}m_{0}\sqrt{1-g_{tt}\dot{t}_{1}\dot{t}_{2}-g_{rr}\dot{r}_{1}\dot{r}_{2}-g_{\phi\phi}\dot{\phi}_{1}\dot{\phi}_{2}},
\end{equation}
and obtain CM energy as the following,
\begin{equation}\label{s17}
\tilde{E}_{CM}=1+\frac{{\mathcal{H}}^{2/3}}{f}E_{1}E_{2}-\frac{R^{2}}{r^{2}}{\mathcal{H}}^{-1/3}L_{1}L_{2}-G_{1}G_{2},
\end{equation}
where,
\begin{eqnarray}\label{s18}
G_{1}&=&\sqrt{1+\frac{{\mathcal{H}}^{2/3}}{f}E_{1}^{2}-\frac{R^{2}}{r^{2}}{\mathcal{H}}^{-1/3}L_{1}^{2}},\nonumber\\
G_{2}&=&\sqrt{1+\frac{{\mathcal{H}}^{2/3}}{f}E_{2}^{2}-\frac{R^{2}}{r^{2}}{\mathcal{H}}^{-1/3}L_{2}^{2}},
\end{eqnarray}
and we used the following re-scaling,
\begin{equation}\label{s19}
\tilde{E}_{CM}\equiv(\frac{E_{CM}}{\sqrt{2}m_{0}})^{2}.
\end{equation}
It is clear that near the black hole ($r\rightarrow r_{+}$) we have
$f\rightarrow0$ and CM energy will be infinite as expected.
\section{Uncharged black hole}
In order to find effect of black hole charge on the CM energy, first
we study uncharged black hole. If we set $q_{i}=0$, then
${\mathcal{H}}=1$ and CM energy will be,
\begin{equation}\label{s20}
\tilde{E}_{CM}=1+\frac{1}{f_{0}}E_{1}E_{2}-\frac{R^{2}}{r^{2}}L_{1}L_{2}-G_{01}G_{02},
\end{equation}
where,
\begin{eqnarray}\label{s21}
f_{0}&=&1-\frac{\mu}{r^{2}}+\frac{r^{2}}{R^{2}},\nonumber\\
G_{01}&=&\sqrt{1+\frac{1}{f_{0}}E_{1}^{2}-\frac{R^{2}}{r^{2}}L_{1}^{2}},\nonumber\\
G_{02}&=&\sqrt{1+\frac{1}{f_{0}}E_{2}^{2}-\frac{R^{2}}{r^{2}}L_{2}^{2}},
\end{eqnarray}
From the Fig. 1 we can see that the maximum energy is near the black
hole horizon. Choosing $R=1$ and $\mu=1$ tells that $r_{+} =
0.78615$ which yields to $\tilde{E}_{CM}=2.5$, while the maximum of
CM energy is about 4.8. Also we see negative energy in the Fig. 1
which shows regions far from the black hole. However at
$r\rightarrow\infty$, as we expected, the CM energy yields to zero.

\begin{figure}[th]
\begin{center}
\includegraphics[scale=.25]{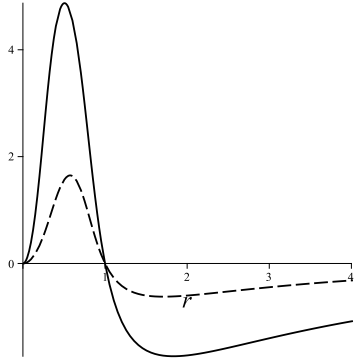}
\caption{Center of mass energy of uncharged black hole with $\mu=1$,
$R=1$. We choose $L_{1}=L_{2}=1$ and $E_{1}=1$. Then solid line
drawn for $E_{2}=10$ and dashed line for $E_{2}=5$.}
\end{center}
\end{figure}

\section{Three-charged black hole}
In that case we set $q_{i}=q$ and find that the black hole charge
restricted to have positive CM energy near the black hole. By
choosing $L_{1}=L_{2}=1$, $E_{1}=1$ and $E_{2}=10$ we obtain $q<0.6$
is necessary to have positive energy near the black hole horizon. In
that case we find that the black hole charge may be decreased or
increased the CM energy. This point illustrated in the Fig. 3.

\begin{figure}[th]
\begin{center}
\includegraphics[scale=.25]{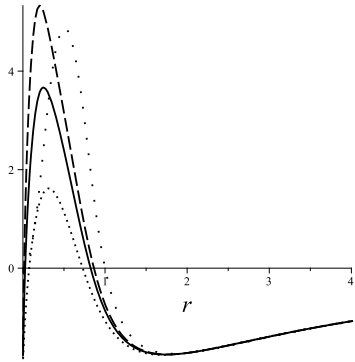}
\caption{Center of mass energy of three-charged black hole with
$\mu=1$, $R=1$. We choose $L_{1}=L_{2}=1$, $E_{1}=1$ and $E_{2}=10$.
Then, space-dotted line drawn for $q=0$, dashed line drawn for
$q=0.24$, solid line drawn for $q=0.3$ and dotted line drawn for
$q=0.42$.}
\end{center}
\end{figure}

\begin{figure}[th]
\begin{center}
\includegraphics[scale=.25]{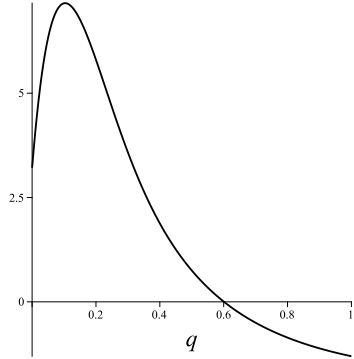}
\caption{Center of mass energy versus black hole charge with
$\mu=1$, $R=1$. We choose $L_{1}=L_{2}=1$, $E_{1}=1$ and
$E_{2}=10$.}
\end{center}
\end{figure}

\section{Conclusion}
In this paper we considered STU black hole and calculate center of
mass energy of two colliding particles near the black hole horizon.
We found that the black hole charge may be increased or decreased
the CM energy which depend on distance from the black hole. Also it
is clear that the CM energy on the black hole horizon will be
infinite.\\
We found that far from the black hole the CM energy yields to zero
and black hole charge is not important parameter.\\
For the future work and complete this job it is interesting to
discuss about CM energy in extremal limits of STU black hole, also
one can consider this black hole in the flat space.

\end{document}